\documentclass[conference]{IEEEtran}
\IEEEoverridecommandlockouts

\usepackage{cite}
\usepackage{algorithmic}
\usepackage{textcomp}
\usepackage{amsmath,amssymb,amsfonts}
\usepackage{algorithm}
\usepackage{array}
\usepackage[caption=false,font=normalsize,labelfont=sf,textfont=sf]{subfig}
\usepackage{stfloats}
\usepackage{url}
\usepackage{verbatim}
\usepackage{graphicx}
\usepackage{multirow}
\usepackage{booktabs}
\usepackage{colortbl}
\usepackage[table]{xcolor}

\def\BibTeX{{\rm B\kern-.05em{\sc i\kern-.025em b}\kern-.08em
    T\kern-.1667em\lower.7ex\hbox{E}\kern-.125emX}}
\begin{document}

\title{SpikEmo: Enhancing Emotion Recognition With
Spiking Temporal Dynamics in Conversations}

\author{
    \IEEEauthorblockN{1\textsuperscript{st} Xiaomin Yu \IEEEauthorrefmark{1}\thanks{\IEEEauthorrefmark{1}Equal Contribution}}
    \IEEEauthorblockA{\textit{Great Bay University}\\
    Guangdong, China \\
    yuxm02@gmail.com}
    \and
    \IEEEauthorblockN{2\textsuperscript{st} Feiyang Wang\textsuperscript{*}}
    \IEEEauthorblockA{\textit{Central South University} \\
    Changsha, Hunan, China \\
8207210931@csu.edu.cn}
    \and
    \IEEEauthorblockN{3\textsuperscript{nd} Ziyue Qiao\IEEEauthorrefmark{2}\thanks{\IEEEauthorrefmark{2}Corresponding Author}}
    \IEEEauthorblockA{\textit{Great Bay University}\\
    Guangdong, China \\
    ziyuejoe@gmail.com}
}

\maketitle

\def\method{SpikEmo}

\begin{abstract}
In affective computing, the task of Emotion Recognition in Conversations (ERC) has emerged as a focal area of research. The primary objective of this task is to predict emotional states within conversations by analyzing multimodal data including text, audio, and video. While existing studies have progressed in extracting and fusing representations from multimodal data, they often overlook the temporal dynamics in the data during conversations. To address this challenge, we have developed the \method{} framework, which is based on spiking neurons and employs a Semantic \& Dynamic Two-stage Modeling approach to more precisely capture the complex temporal features of multimodal emotional data. Additionally, to tackle the class imbalance and emotional semantic similarity problems in the ERC tasks, we have devised an innovative combination of loss functions that significantly enhances the model's performance when dealing with ERC data characterized by long-tail distributions. Extensive experiments conducted on multiple ERC benchmark datasets demonstrate that \method{} significantly outperforms existing state-of-the-art methods in ERC tasks. Our code is available at \url{https://github.com/Yu-xm/SpikEmo.git}.
\end{abstract}


\section{Introduction}

Emotion Recognition in Conversations (ERC) task \cite{b1,b2,b3,b4,b5,b6,b7,b8,b9,b10,b11,b12} aims to recognize emotions in each utterance based on text, audio, and visual information from the speaker, which is crucial for several applications such as Human-Computer Interaction and Mental Health Analysis.
In conversations, emotional information is presented through text, audio, and visual modalities, exhibiting complex dynamic characteristics over time \cite{b13,b14}. Thus, accurately capturing the temporal features in multimodal data is crucial for developing effective emotion recognition models. While existing studies have progressed in extracting and fusing representations from multimodal data, they often overlook the temporal dynamics in the data during conversations. 

\begin{figure}[ht]
  \centering
  \includegraphics[width=0.45\textwidth]{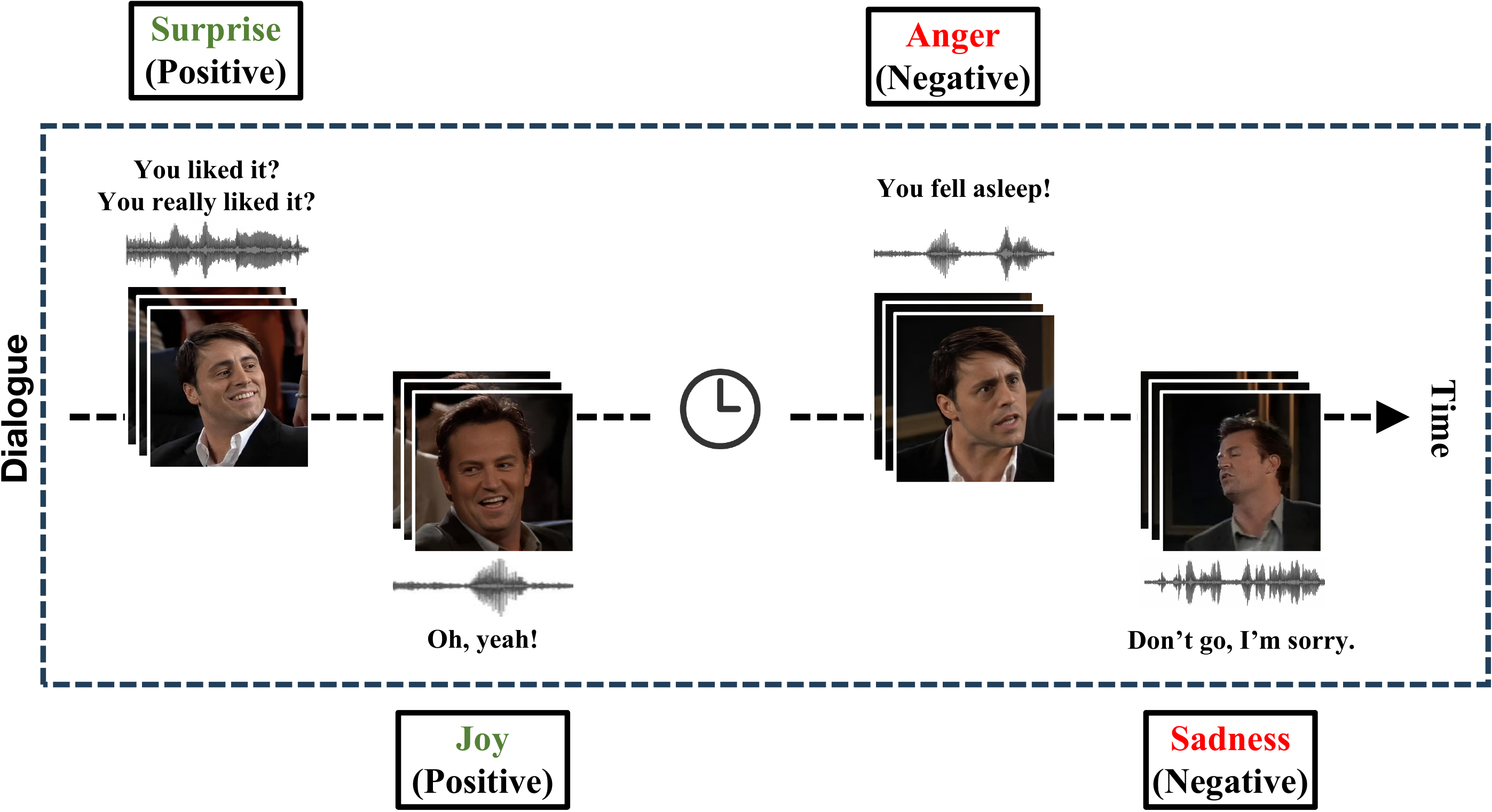} 
  \caption{An example of the ERC task from the MELD dataset.}
  \label{fig:intro} 
\end{figure}

To model the temporal information in the data, we propose a two stage modality-temporal modeling approach. In the first stage at the modality level, we employ targeted feature extraction models for different modality data characteristics to obtain modal representations. In the second stage at the temporal level, we introduce a feature-level dynamic contextualized modeling module, which combines the sequence modeling capabilities of the Transformer with the temporal dynamics handling of Spiking Neural Networks (SNNs) \cite{b15,b16,b17}, making it possible to effectively capture the complex temporal information in text, audio, and visual data.  The essence of the feature-level dynamic contextualized modeling module is the spiking behavior of pulse neurons, which mimics the temporal dynamic characteristics of biological neurons. This dynamic nature allows the feature-level dynamic contextualized modeling module to capture changes in input data at different times, highlighting moments particularly important for emotion state recognition.

\begin{figure}[h]
    \centering
    \includegraphics[width=0.46\textwidth]{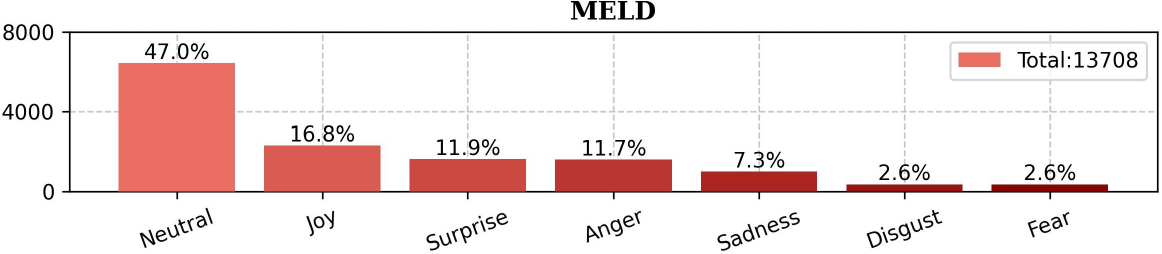}

    \vspace{1mm}
    
    \includegraphics[width=0.46\textwidth]{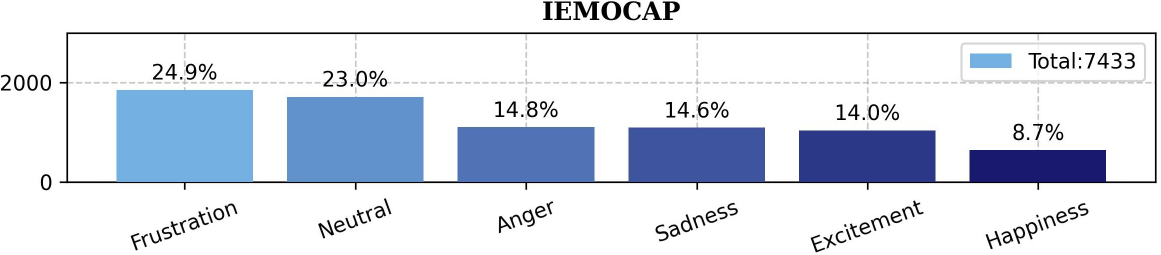}

    \caption{The data distribution of MELD and IEMOCAP. Both datasets exhibit a noticeable long-tail distribution.}
    \label{fig:MELD_dis}
\end{figure}

In addition, there are two inherent problems in the ERC dataset \cite{b18,b19}. First of all, In the ERC dataset, it is challenging to accurately classify certain semantically \cite{b20,b21} related emotions such as disgust and anger. Although these emotions show subtle differences at the literal level, they exhibit significant similarities in human cognitive, emotional, and physiological characteristics, making their differentiation particularly complex in ERC tasks. Furthermore, natural tendencies in emotional expression, sociocultural factors, and data collection biases contribute to a long-tail distribution \cite{b22,b23,b24,b25,b26} characteristic in ERC datasets. This is manifested as frequent occurrences of common emotions in everyday conversations within the dataset, while some emotions are rare. The data distribution shown in Fig. \ref{fig:MELD_dis} illustrates this point well. The sample imbalance inherent in ERC tasks \cite{b22} leads to models that are prone to overfitting common emotions while struggling to accurately recognize rarer emotional categories. To address this issue, we employ a class-balancing loss that assigns higher weights to difficult-to-classify samples, enhancing the model's focus on these samples.

In summary, we propose the \method{} framework to address the core challenges in the ERC task. This framework effectively handles the temporal dynamics and semantic complexity in dialogue, processing the rich variations in emotions. Experimental validation on standard datasets such as MELD\cite{b18} and IEMOCAP\cite{b19} demonstrates the potential of this approach in accurately recognizing and understanding diverse emotional states. The main contributions of this paper can be summarized in three areas:

\begin{itemize}
  \item We introduce a new framework called \method{} tailored for the ERC task, proposing a two-phase modal-temporal modeling approach to represent different modal representations and extract spiking temporal features.
  \item We utilize state-of-the-art modal fusion techniques and design a combined optimization objective for the ERC task to address the issues of class imbalance and semantic similarity between different emotions.
  \item We conducted extensive experiments on the standard ERC task datasets, MELD and IEMOCAP. Our method surpasses the existing state-of-the-art methods based on discriminative models and LLMs, confirming the efficacy of our model.
\end{itemize}

\label{sec:intro}

\section{Related works}

\textbf{Emotion Recognition in Conversation (ERC)} \cite{b1, b2, b3, b4, b5, b6, b7, b8, b9, b10, b11, b12, b13, b14} aims to identify the emotional expressions in human dialogues such as happiness, sadness, anger, surprise, etc \cite{b18,b19}. It requires a comprehensive understanding of each participant's emotional attitude based on information sources from multiple modalities, including textual content of dialogues, audio for vocal emotions, and visual cues for facial expressions. Recent advancements in pre-trained large language models (LLMs) \cite{b31} have demonstrated exceptional performance in natural language processing tasks. Based on these developments, some studies have adopted a generative architecture based on LLMs for Emotion Recognition in Conversation (ERC) tasks. The notable improvement in performance with this approach has validated the potential of LLMs in this domain. \cite{b30} However, the effectiveness of LLMs heavily relies on substantial computational resources and extensive training data, which may not be feasible in resource-constrained environments. Therefore, continuing to explore the viability of traditional discriminative models for efficiently handling ERC tasks remains a critical area of research. Additionally, certain intense emotions being rare in everyday conversations because of their specific occurrence conditions, and the ERC datasets consequently still an issue of class imbalance \cite{b22, b23, b24, b25, b26}.

\textbf{Spiking Neural Networks (SNNs)} Spiking Neural Networks (SNNs) are a novel type of neural network model, characterized by their use of discrete spike sequences for computation and information transmission \cite{b15}. Inspired by the biological neural systems, these networks communicate through mechanisms that emulate the pulse transmissions of biological neurons. In SNNs, neurons convert continuous input values into spike sequences, efficiently encoding and responding to temporal changes through their spiking activity. This time-based signal processing capability enables SNNs to naturally adapt to the dynamics of data, especially excelling in handling time-related data. In recent years, researchers have begun to explore the application of the Transformer model to SNNs. The Transformer model, having achieved significant success in various natural language processing and computer vision tasks, has been integrated with SNNs \cite{b16, b17}. The first spiking Transformer model introduced spiking self-attention mechanisms to model visual features using sparse Query, Key, and Value matrices. Additionally, some studies have proposed variants using spatio-temporal attention to better incorporate the attention mechanism within the Transformer. These innovations open up potential for the application of pulsed attention mechanisms in complex tasks across natural language processing and computer vision.

\section{\textbf{Method}}

\begin{figure*}[h]
  \centering
  \includegraphics[width=0.96\textwidth]{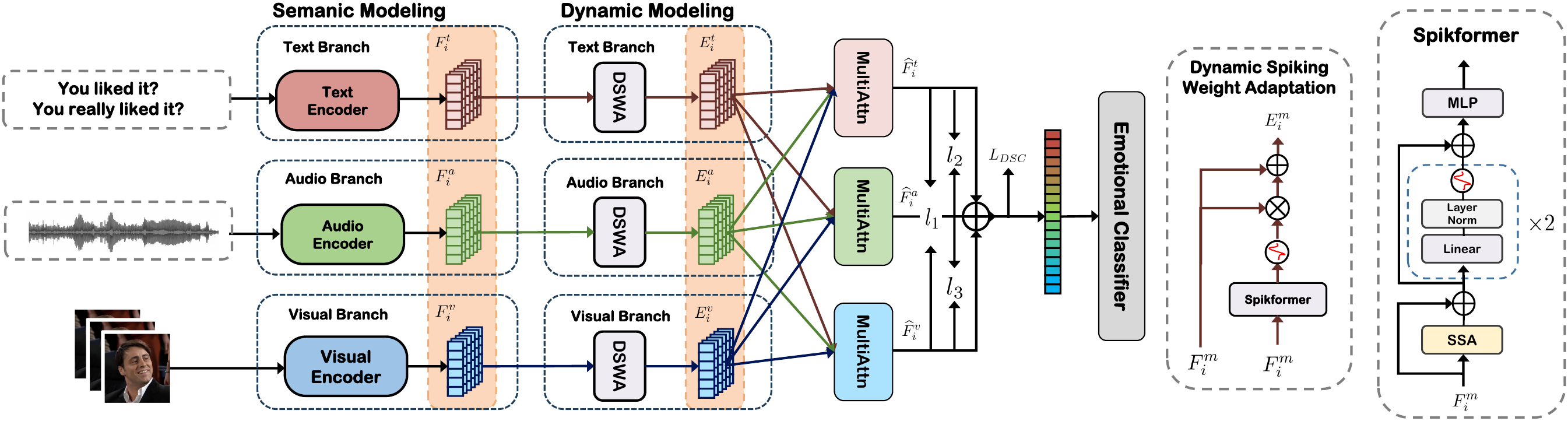} 
  \caption{Overall framework of \method{}. The modality level Semantic modeling extracts the contextualized modality representations, the feature level dynamic contextualized modeling extracts cross-modality temporal information, and the $L_{corr}$ and $L_{DSC}$ losses are proposed to capture correlations and avoid the long-tail problem in model training.}
  \label{fig:intro} 
\end{figure*}

\subsection{\textbf{Semantic \& Dynamic Two-stage Modeling}}

In the task of ERC, we consider a set of dialogues $D = \left\{\left(s_1, U_1\right), \left(s_2, U_2\right), ..., \left(s_k,U_k\right)\right\}$, where  $k$ is the total number of turns in the dialogue, $s_i$ is the speaker of the $i$-th turn, and $U_i$ is the corresponding utterance. Each utterance $U_i$ includes three modalities---$U_i = \{U^t_i, U^a_i, U^v_i\}$,  where $U^t_i, U^a_i, U^v_i$ represent the textual, audio, and visual sequences of the utterance $U_i$, respectively.
Each utterance $U_i$ is associated with an emotional label $y_i \in \mathcal{C}$, where $\mathcal{C}$ is a predefined set of emotional categories. 
The task is to use the combined multimodal information of $U_i$ and its context information in $D$ to predict its emotional label $y_i$.

\subsubsection{\textbf{Modality level Semantic modeling}}
In this section, we extract modality features for text, audio, and video modalities. Specifically, for text modality, given the textual utterance sequences $\{U_1^t,U_2^t,...,U_k^t\}$, we concatenate the sequences together and employ the pre-trained RoBERTa \cite{b27} to process it into textual representations---$\{F_1^t,F_2^t,...,F_k^t\}= \text{RoBERTa}(\{U_1^t,U_2^t,...,U_k^t\})$, 
Similarly, for audio modality, we employ OpenSMILE \cite{b28} to extract 6373-dimensional features for each utterance audio. 
Then, we employ DialogueRNN\cite{b29} to capture contextualized audio representations $\{F_1^a,F_2^a,...,F_k^a\}$ for each audio clip.
For visual modality, we employ VisExtNet\cite{b11} to extract facial expression features of interlocutors from multiple frames,  excluding redundant scene-related information
Subsequently, DialogueRNN\cite{b29} is also used to learn contextualized visual representations $\{F_1^v,F_2^v,...,F_k^v\}$ for each video frame.
Since the processed representation sequences $F_i^t,F_i^a,F_i^v$ for textual, audio, and visual modalities may vary in sequence lengths and representation dimensions, we apply masking and use linear layers to transform these sequences into a uniform shape, $\mathbb{R}^{l*h}$, for subsequent unified processing.



\subsubsection{\textbf{Feature level dynamic contextualized modeling}}

We introduce the dynamic spiking weight adaptation (DSWA) module to extract temporal features of inputs. The spiking self-attention layer is the core of DSWA, utilizing the dynamics of SNNs to simulate dependencies and interactions over time. In SSA, the self-attention mechanism is adjusted to accommodate the characteristics of spiking neurons, thus more effectively capturing the dynamic changes in time-series data. For each modality $m\in\{t, a, v\}$, we first compute its queries $Q^m_{i}$, keys $K^m_{i}$, and values $V^m_{i}$, which are obtained by transforming the original modality feature $F^m_i$ through linear layers within the Spikformer. The SSA layer can be represented as:


\begin{align}
\text{SSA}(Q^m_{s}, K^m_{s}, V^m_{s}) = S(\tau \cdot Q^m_{s}(K^m_{s})^T)V^m_{s}),
\end{align}



\noindent where $S$ is the step function, simulating the firing behavior of spiking neurons; $\tau$ is a scaling factor used to adjust the intensity of the spike response. This process captures dependencies between different time points and encodes them into spike signals.

By stacking multiple SSA layers followed by Linear and Normalization layers, we obtain the output spiking representation $E^m_i$,  we use the softmax function to determine the importance weights for each modality feature. Then, We multiply these weights with the original modality feature to highlight those moments that are particularly important for recognizing the emotional state. Finally, we introduce a residual structure to merge the time-weighted enhanced features with the original modality features. The final modality features can be written as follows:
\begin{align}
F^m_i = \delta (E^m_i) \cdot F^m_i + F^m_i, 
\end{align}

\noindent where $\delta(\cdot)$ represents softmax function.

\subsection{\textbf{Heterogeneous Modality Feature Fusion}}

Then, We employ a multimodal fusion network called MultiAttn \cite{b11} to integrate the characteristics of these different modalities. MultiAttn utilizes a bidirectional multi-head cross-attention mechanism. A complete MultiAttn architecture comprises three main components: $MultiAttn_{text}$, $MultiAttn_{audio}$, and $MultiAttn_{visual}$, each consisting of $T$ layers, where the output of first stage attention layer is fed into the subsequent layer as the new Query. For each modalities $x$, $y$, and $z$, where $x$, $y$, and $z$ each correspond to one of the modalities in $m\in\{t,a,v\}$. The first stage of the attention mechanism can be written as follows:

\begin{equation}
  F^{xy,(l)}_i = \text{MultiAtten}(F^{x,(l-1)}_i, F^{y,(l-1)}_i, F^{y,(l-1)}_i)
\end{equation}








\noindent where $l$ is the index of the layer number and $F^{x,(0)}_i = F^{x}_i$, $\text{MultiAtten}$ represents the multi-head self-attention layer.




Then, the output from the first stage is utilized as the new Query, which can be written as follows:

\begin{equation}
  F^{xyz,(l)}_i = \text{MultiAtten}(F^{xy,(l-1)}_i, F^{z,(l-1)}_i, F^{z,(l-1)}_i).
\end{equation}






By stacking multiple \textit{MultiAtten} layers followed by Linear layers for textual, audio, and visual modalities, the final representations $\widehat F^{t}_i, \widehat F^{a}_i, \widehat F^{c}_i$ for each utterance is obtained.

\subsection{\textbf{Optimization Object}}

To enhance the performance of the model on the ERC task, we considered two critical issues in the optimization process: (1) In ERC tasks, certain emotions (such as anger and disgust) exhibit highly similar semantic expressions in multimodal contexts; (2) As shown in the figure, the datasets for ERC tasks often exhibit a severe long-tail distribution problem.

To solve the problem (1), we introduce the $L_{corr}$. Specifically, we use $L_{corr}$ to effectively capture and utilize the complex correlations between different Modality features. $L_{corr}$ consists of three parts:


\vspace{-6pt}

\begin{equation}
\begin{aligned}
l_1 = -\mathbb{E}_{X,Y \sim D} [(\widehat F_{x})^T \widehat F_{y}] + \frac{1}{2} \text{tr}[cov(\widehat F_{x}) cov(\widehat F_{y})]
\end{aligned}
\end{equation}

\vspace{-12pt}

\begin{equation}
\begin{aligned}
l_2 = -\mathbb{E}_{X,Z \sim D} [(\widehat F_{x})^T \widehat F_{z}] + \frac{1}{2} \text{tr}[cov(\widehat F_{x})cov(\widehat F_{z})] 
\end{aligned}
\end{equation}

\vspace{-12pt}

\begin{equation}
\begin{aligned}
l_3 = -\mathbb{E}_{Y,Z \sim D} [(\widehat F_{y})^T \widehat F_{z}] + \frac{1}{2} \text{tr}[cov(\widehat F_{y})cov(\widehat F_{z})]
\end{aligned}
\end{equation}

\noindent where each row in Equations (5), (6), and (7) uses the HGR maximal correlation implemented by deep learning to compute the correlation between \(\widehat F_{x}\) and \(\widehat F_{y}\), \(\widehat F_{x}\) and \(\widehat F_{z}\), as well as \(\widehat F_{y}\) and \(\widehat  F_{z}\) on the complete data.


\begin{equation}
\begin{aligned}
L_{corr} = \alpha_1 \cdot l_1 + \alpha_2 \cdot l_2 + \alpha_3 \cdot l_3,
\end{aligned}
\end{equation}

\noindent where $\alpha_1 , \alpha_2, \alpha_3$ are the loss wrights.

To solve the problem (2), we introduce DSC Loss \cite{b22}. The DSC loss adapts a self-regulating mechanism that reduces the focus on easily predictable samples (i.e., samples with prediction probabilities close to 1 or 0) by using $1 - p$ as a scaling factor. DSC loss achieves more balanced model optimization when dealing with imbalanced datasets. DSC Loss can be defined as follows:


\begin{align}
L_{\text{DSC}} &= \frac{1}{N} \sum_{i=1}^{N} \left[ 1 - \frac{2 \cdot (1 - p_{i,c})^\alpha \cdot p_{i,c} \cdot y_{i,c} + \gamma}{(1 - p_{i,c})^\alpha \cdot p_{i,c} + y_{i,c} + \gamma} \right], 
\end{align}

\noindent where $y_{i,c}$ is the ground-truth label of the $c$-th sentence in the $i$-th dialogue. $\alpha$ is a weighting factor used to enhance the impact of tail data, and $\gamma$ is a smoothing parameter.

Then, we adopted the cross-entropy (CE) loss function to reduce the discrepancy between the real and predicted labels.  CE loss can be written as follows:
\begin{align}
L_{\text{CE}} = -\frac{1}{N} \sum_{i=1}^{N} \sum_{c=1}^{C} y_{i,c} \log p_{i,c},
\end{align}

\noindent where $C$ is the number of emotional category.

Finally, the overall optimization objective can be written as follows:




\begin{align}
L = \min_{\theta}\left\{ L_{\text{CE}} + \lambda_1 \cdot L_{\text{DSC}} + \lambda_2 \cdot L_{\text{corr}} \right\},
\end{align}

\noindent where $\lambda_1$, $\lambda_2$ are the loss wrights, and $\theta$ is overall parameters.

\subsection{\textbf{Final Emotional Classification}}

\noindent In conclusion, the final emotional labels are obtained from the following formula:
\begin{align}
\mathbf{p}_i &= \delta(\phi_{\text{MLP}}([\widehat F^{t}_i; \widehat F^{a}_i; \widehat F^{c}_i])),
\end{align}
\begin{align}
\hat{\mathbf{y}}_i = \arg\max_{t}(\mathbf{p}_i[t]),
\end{align}

\noindent where $p_i[t]$ represents the probability of the $t$-th category.

\section{\textbf{Experiments}}

\subsection{\textbf{Setup}}

\begin{table}[ht]
\centering
\renewcommand{\arraystretch}{1.}
\caption{Statistical Data for the MELD and IEMOCAP Datasets. Note: T represents Text, A represents Audio, and V represents Video.}
\label{table1:statistics}
\resizebox{0.98\columnwidth}{!}{
\begin{tabular}{@{}c|ccc|ccc|cc@{}}
\toprule[1.5pt]
\multirow{3}{*}{Datasets} & \multicolumn{3}{c|}{Dialogue} & \multicolumn{3}{c|}{Utterances} & \multicolumn{2}{c}{Attributes} \\ 
\cmidrule(lr){2-4} \cmidrule(lr){5-7} \cmidrule(lr){8-9}
& Train & Valid & Test & Train & Valid & Test & Modality & Classes \\ 
\midrule
MELD   & 1038 & 114 & 280 & 9989 & 1109 & 2610 & T, A, V & 6 \\
IEMOCAP & 108 & 12 & 31 & 5163 & 647 & 1623 & T, A, V & 7 \\
\bottomrule[1.5pt]
\end{tabular}
}
\end{table}

\subsubsection{\textbf{Datasets}}

To verify the model’s generality, we conducted experiments using the widely recognized standard datasets MELD and IEMOCAP. The statistical overview of the related datasets is presented in Table \ref{table1:statistics}. 

The \textbf{MELD} dataset, derived from the TV series "Friends", is a multimodal and multiparty dataset focusing on emotion recognition from raw dialogue transcripts, audio, and video. It features over 13,000 utterances and 1,400 conversations annotated with one of seven emotional categories: anger, disgust, fear, happiness, neutrality, sadness, and surprise. 

The \textbf{IEMOCAP} dataset is a comprehensive multimodal dataset developed by the SAIL Lab at the University of Southern California. It consists of approximately 12 hours of audiovisual data involving ten actors who engage in improvised dyadic interactions. These scenarios are specifically crafted to elicit a variety of emotional states such as happiness, sadness, anger, excitement, frustration, and neutral.

\subsubsection{\textbf{Discriminative models Baselines}} Discriminative models for ERC tasks, focus on directly mapping multimodal emotional data inputs to output sentiment labels. The defining characteristic of these models is their capacity to achieve promising results even with limited training data. Typically, they leverage various advanced architectures to capture contextual and speaker-specific information effectively, utilizing techniques such as recurrent neural networks \cite{b1,b2,b4}, graph-based models \cite{b3,b6,b8,b10,b12}, and transformer variants \cite{b5,b11,b12}. These approaches often involve modeling dialogue context \cite{b1,b2,b3,b8}, speaker states\cite{b2}, and cross-modal interactions \cite{b3,b8,b10,b11} through methods like graph convolution\cite{b3}, positional encoding\cite{b6}, multimodal fusion\cite{b7,b8,b9,b10,b11}, and quantum-inspired\cite{b7} frameworks.

\subsubsection{\textbf{LLMs-based Baselines.}} LLMs-based models, through pre-training on extensive datasets, capture a wealth of linguistic features and contextual information. Recent studies \cite{b30} have highlighted the advantages of these models in ERC tasks. A key advantage of these models is their ability to generalize based on the extensive prior knowledge accumulated during the pre-training process, thereby enhancing performance on target tasks.

\subsubsection{\textbf{Implementation Details}} For the MELD dataset, we utilize a batch size of 64 and assign the loss weights $\lambda_1$, $\lambda_2$, and $\lambda_3$ as 0.3, 1, and 0.4 respectively. For the IEMOCAP dataset, the batch size is set at 32, with loss weights $\lambda_1$, $\lambda_2$, and $\lambda_3$ configured to 0.4, 0.6, and 1, respectively. Regarding generic settings applicable across all datasets, the MultiAttn model consistently employs 6 layers, whereas the Spikformer model uses 2 layers with a time step of T=32. The parameters $\alpha$ and $\gamma$ for the $L_{corr}$ loss function are set to 1.5 and 0.5, respectively. Training is conducted over 50 epochs using the Adam optimizer with $\beta_1$ = 0.9 and $\beta_2$ = 0.99; the initial learning rate is set at 0.0001, with a decay rate of 0.95 every 10 epochs. Additionally, the L2 regularization weight is set at 0.00001.

\subsubsection{\textbf{Metrix}} We utilized the \textbf{Weighted F1} score as the evaluation metric in our experiments.

\subsection{\textbf{Main Result}}

\begin{table*}[ht]
\centering
\setlength{\tabcolsep}{2pt}
\renewcommand{\arraystretch}{1.56}
\caption{Comparative Results on the MELD and IEMOCAP datasets.}
\label{table2: main_results}
\resizebox{1.98\columnwidth}{!}{
\begin{tabular}{c|cccccccc|ccccccc}
\toprule[2pt]
\multirow{2}{*}{Models} & \multicolumn{8}{c|}{\textbf{MELD}} & \multicolumn{7}{c}{\textbf{IEMOCAP}} \\
& Neutral & Surprise & Fear & Sadness & Joy & Disgust & Angry & \textbf{W-F1} & Happiness & Sadness & Neutral & Anger & Excitement & Frustration & \textbf{W-F1} \\
\midrule
BC-LSTM \cite{b1} & 73.80 & 47.70 & 5.40 & 25.10 & 51.30 & 5.20 & 38.40 & 55.90 & 34.43 & 60.87 & 51.81 & 56.73 & 57.95 & 58.92 & 54.95 \\
DialogueRNN \cite{b2} & 76.23 & 49.59 & 0.00 & 26.33 & 54.55 & 0.81 & 46.76 & 58.73 & 33.18 & 78.80 & 59.21 & 65.28 & 71.86 & 58.91 & 62.75 \\
DialogueGCN \cite{b3} & 76.02 & 46.37 & 0.98 & 24.32 & 53.62 & 1.22 & 43.03 & 57.52 & 51.87 & 76.76 & 56.76 & 62.26 & 72.71 & 58.04 & 63.16 \\
IterativeERC \cite{b4} & 77.52 & 53.65 & 3.31 & 23.62 & 56.63 & 19.38 & 48.88 & 60.72 & 53.17 & 77.19 & 61.31 & 61.45 & 69.23 & 60.92 & 64.37 \\
CESTa \cite{b5} & - & - & - & - & - & - & - & 58.36 & 47.70 & 80.82 & 64.76 & 63.41 & 75.95 & 62.65 & 67.10 \\
RGAT-ERC \cite{b6} & - & - & - & - & - & - & - & 60.91 & 51.62 & 77.32 & 65.42 & 63.01 & 67.95 & 61.23 & 65.22 \\
QMNN \cite{b7} & 77.00 & 49.76 & 0.00 & 16.50 & 52.08 & 0.00 & 43.17 & 58.00 & 39.71 & 68.30 & 55.29 & 62.58 & 66.71 & 62.19 & 59.88 \\
MMGCN \cite{b8} & - & - & - & - & - & - & - & 58.65 & 42.34 & 78.67 & 61.73 & 69.00 & 74.33 & 62.32 & 66.22 \\
MVN \cite{b9} & 76.65 & 53.18 & 11.70 & 21.82 & 53.62 & 21.86 & 42.55 & 59.03 & \underline{55.75} & 73.30 & 61.88 & 65.96 & 69.50 & 64.21 & 65.44 \\
MM-DFN \cite{b10} & 77.76 & 50.69 & - & 22.93 & 54.78 & - & 47.82 & 59.46 & 42.22 & 78.98 & 66.42 & \underline{69.77} & 75.56 & \underline{66.33} & 68.18 \\
MultiEMO \cite{b11} & \underline{78.22} & \underline{56.52} & \underline{21.78} & \underline{39.14} & \underline{62.02} & \underline{23.76} & \underline{52.91} & \underline{64.43} & 52.55 & \underline{83.44} & 66.42 & 66.07 & 74.02 & 63.04 & 68.33 \\
GA2MIF \cite{b12}& 76.92 & 49.08 & - & 27.18 & 51.87 & - & 48.52 & 58.94 & 46.15 & \textbf{84.50} & \underline{68.38} & \textbf{70.29} & \underline{75.99} & \textbf{66.49} & \underline{70.00} \\
\midrule
\textbf{\method{}} & \textbf{79.53} & \textbf{58.40} & \textbf{21.95} & \textbf{40.00} & \textbf{63.89} & \textbf{28.85} & \textbf{54.20} & \textbf{65.92} & \textbf{58.16} & 82.17 & \textbf{72.61} & 66.67 & \textbf{79.01} & 64.82 & \textbf{71.50} \\
\bottomrule[2pt]
\end{tabular}
}
\end{table*}

We systematically compare the performance of our model against two categories of baselines: traditional discriminative methods and those based on large language models (LLMs).

\textbf{Compare with discriminative models Baselines.} We compared our approach with the existing SOTA methods on the MELD and IEMOCAP, as shown in TABLE \ref{table2:  main_results}. On the MELD dataset, our method improved by \textbf{1.49\%} compared to the previous best model; on the IEMOCAP dataset, it improved by \textbf{1.50\%}. Moreover, in tail categories, such as the Fear category in the MELD dataset and the Happiness category in the IEMOCAP dataset, our model outperformed the previous best models by \textbf{0.17\%} and \textbf{2.41\%}, respectively, further validating the effectiveness of our model.

\textbf{Compared with LLM-based Baselines.} We evaluated our model against state-of-the-art Large Language Model (LLM) baselines. As outlined in TABLE \ref{table3:LLMs-based}, our model features significantly fewer parameters. While our scores may not be the highest, the efficiency of our model is markedly superior given the reduced complexity. Specifically, our model has only \textbf{0.5\%} of the parameter count of the leading LLMs but achieves commendably close performance metrics. This demonstrates our model’s enhanced efficiency and underscores its potential for applications requiring lower computational resources and energy consumption.



\begin{table}[h]
\centering
\renewcommand{\arraystretch}{1.2}
\caption{comparison results with LLM-based models on the MELD and IEMOCAP datasets.}
\begin{tabular}{c|cc|c}
\toprule[1.6pt]
\textbf{Models} & \textbf{IEMOCAP} & \textbf{MELD} & \textbf{Para} \\ 
\midrule
\multicolumn{4}{c}{\textbf{LoRA + Backbone}} \\ \midrule
ChatGLM & 17.98 & 40.54 & \multirow{4}{*}{$\geq$ 6B} \\
ChatGLM2 & 52.88 & 64.85 &  \\
Llama & 55.81 & 66.15 &  \\
Llama2 & 55.96 & 65.84 &  \\ 
\midrule
\multicolumn{4}{c}{\textbf{LoRA + InstructERC}} \\ \midrule
ChatGLM & 36.04 & 46.41 & \multirow{4}{*}{$\geq$ 6B} \\
ChatGLM2 & 67.54 & 65.58 &  \\
Llama & 64.17 & 67.62 &  \\
Llama2 & \underline{71.39} & \textbf{69.15} &  \\ 
\midrule
\multicolumn{4}{c}{\textbf{Ours (Discriminative model)}} \\ 
\midrule
\textbf{SpikEmo} & \textbf{71.50} & \underline{65.92} & \textbf{31M} \\
\bottomrule[1.6pt]
\end{tabular}
\label{table3:LLMs-based}
\end{table}

\begin{table*}[ht]
\centering
\setlength{\tabcolsep}{5pt}
\renewcommand{\arraystretch}{1.6}
\caption{The results of the ablation study.} \label{table: ablation loss}
\resizebox{\textwidth}{!}{%
\begin{tabular}{c|cccccccc|ccccccc}
  \toprule[2pt]
  \multirow{2}{*}{Model} & \multicolumn{8}{c|}{\textbf{MELD}} & \multicolumn{7}{c}{\textbf{IEMOCAP}} \\
  & Neutral & Surprise & Fear & Sadness & Joy & Disgust & Angry & \textbf{W-F1} 
  & Happiness & Sadness & Neutral & Anger & Excitement & Frustration & \textbf{W-F1} \\
  \midrule
  w/o $\mathcal{L}_{DSC}$ \& w/o $\mathcal{L}_{corr}$
  &74.85 &54.52 &15.91 &\underline{37.68} &58.73 &16.16 &\underline{51.98} &61.53 &52.63 &80.79 &68.04 &66.00 &75.52 &\underline{63.55} &68.71 \\
  w/o $\mathcal{L}_{DSC}$  
  &\underline{79.14} &\underline{58.17} &15.15 &37.34 &\underline{60.85} &\underline{28.04} &48.31 &\underline{64.10} &56.27 &81.97 &\underline{71.38} &\underline{66.00} &76.26 &61.89 &69.75 \\
  w/o $\mathcal{L}_{corr}$  
  &77.72 &57.45 &\underline{19.05} &37.31 &58.80 &26.92 &47.68 &63.00 &\underline{57.34} &\underline{81.97} &69.61 &64.98 &\underline{78.49} &62.40 &\underline{69.84} \\
  w/o $DSWA$ 
  &77.56 &54.21 &\underline{19.28} &33.23 &59.49& \underline{24.74}& \underline{49.70} &62.44& \underline{53.28} &77.21& \underline{70.49}& \underline{65.80}& \underline{77.93} & 62.71 & \underline{69.03} \\
  \midrule
  \textbf{\method{}} & \textbf{79.53} & \textbf{58.40} &\textbf{21.95} & \textbf{40.00} & \textbf{63.89} & \textbf{28.85} & \textbf{54.20} & \textbf{65.92} & \textbf{58.16} & \textbf{82.17} & \textbf{72.61} & \textbf{66.67} & \textbf{79.01} & \textbf{64.82} & \textbf{71.50} \\
  \bottomrule[2pt]
\end{tabular}
}
\end{table*}

\subsection{\textbf{Ablation Study}}

\textbf{Impact of $L_{\text{DSC}}$ \& $L_{\text{corr}}$.} The results are shown in Table \ref{table: ablation loss}. First, we notice that when any component is removed from the model, its performance on all datasets decreases; conversely, integrating all modules results in the best performance of the method. This directly validates the significant contribution of \method{}. Furthermore, when the training process includes $L_{\text{DSC}}$, we clearly observe significant improvements in the performance of \method{} on the MELD and IEMOCAP datasets, especially in the "Fear" category of MELD and the "Happiness" category of IEMOCAP, indicating that $L_{\text{DSC}}$ plays an active role in improving the issue of class imbalance.

\textbf{Impact of Feature Level Dynamic Contextualized Modeling.} To investigate the impact of Feature Level Dynamic Contextualized Modeling, we removed the parts related to it. In the experimental results presented in TABLE \ref{table: ablation loss}. After removing this component, the categorical performance on the MELD and IEMOCAP datasets significantly decreases. This effect is reflected in the decrease of the Weight-F1 scores for both datasets, which dropped by \textbf{3.48\%} and \textbf{2.47\%}, respectively.   

\textbf{Impact of Different Modality Settings.} 
In Table \ref{table5: modality}, we provides a detailed analysis of the model's performance under three types of modality settings: unimodal, bimodal, and trimodal. The findings indicate that the model achieves the best emotion recognition performance when utilizing information from all three modalities (text, audio, and video). This result shows that the integration of multimodal information significantly enhances the accuracy of emotion recognition. It is noteworthy that when the model uses only a single modality as input, the performance based on text data significantly outperforms that based on either audio or video data alone. This phenomenon indicates that the text modality possesses a higher discriminatory ability and informational value in emotion recognition tasks. Further analysis reveals that when the model is trained with combined modalities, its performance remains at a high level as long as the text modality is included. This finding further emphasizes the importance of providing explicit emotional cues in the text, enabling the model to capture the speaker's emotional variations more effectively.

\begin{table}[ht]
\centering
\renewcommand{\arraystretch}{1.2}
\caption{The Results under different modality settings.} 
\label{table5: modality}
\resizebox{0.36\textwidth}{!}{%
\begin{tabular}{c|c|cc}
  \toprule[1.6pt]
  \textbf{Modality} & \textbf{MELD} & \textbf{IEMOCAP} \\
  \midrule
  Audio & 35.91 & 40.67 \\ 
  Visual & 31.27 & 27.44 \\
  Text & 59.56 & 64.77  \\  
  Audio + Visual & 39.98 & 46.80 \\ 
  Text + Audio & 62.22 & \underline{67.06} \\ 
  Text + Visual & \underline{63.22} & 65.15 \\ 
  \textbf{Text + Audio + Visual} & \textbf{65.92} & \textbf{71.50} \\
  \bottomrule[1.6pt]
\end{tabular}
}
\end{table}

\subsection{\textbf{Hyperparameter Experiments}}

\textbf{Impact of Time Constant T.} The time constant \(T\) is a critical parameter in spiking neural networks, determining the response window of spiking neurons to input signals. In spiking neural networks, the firing behavior of neurons (i.e., the generation of spikes) depends not only on the current input but also on the historical input, and the time constant \(T\) is used to modulate this dependency. In the hyperparameter experiments, we conducted an in-depth investigation into the impact of the time constant \(T\) on model performance. Specifically, we set the time constant \(T\) to 2, 4, 8, 16, 32, and 64, respectively, and the experimental results are shown in Fig.4. From the figure, it is clearly observed that as the value of \(T\) increases, the model's performance exhibits a distinct upward trend. This phenomenon indicates that a longer time constant helps better capture the long-term temporal dependencies in the input data, thereby significantly improving the accuracy of emotion recognition.

\begin{figure}[h]
  \centering
  \includegraphics[width=0.46\textwidth]{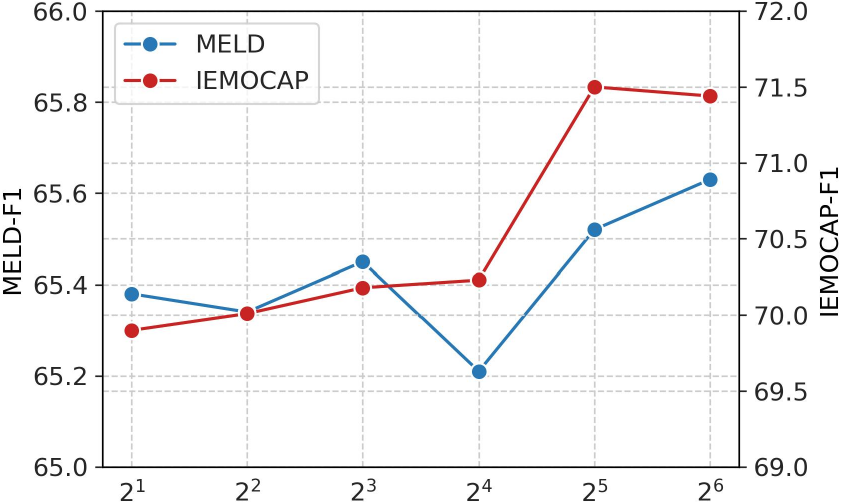} 
  \caption{The impact of T settings on model performance.}
  \label{fig:T} 
\end{figure}

\section{Conclusion}

We proposed the \method{} framework to address key challenges in Emotion Recognition in Conversations (ERC). Our two-stage modality-temporal modeling approach integrates targeted feature extraction and feature-level dynamic contextualized modeling. This enables SpikEmo to effectively capture temporal features and highlight critical emotional transitions.  Our approach also addresses key challenges such as class imbalance and semantic similarity, significantly improving performance on ERC tasks across multiple datasets. Experimental validation on MELD and IEMOCAP demonstrated that SpikEmo surpasses existing state-of-the-art models, including those based on large language models (LLMs), confirming its potential for enhancing emotion recognition in human-computer interaction and mental health analysis.

\end{document}